\documentclass{Interspeech}

\interspeechcameraready 

\title{RA-CLAP: Relation-Augmented Emotional Speaking Style Contrastive Language-Audio Pretraining For Speech Retrieval}

\author[affiliation={1,2}]{Haoqin}{Sun}
\author[affiliation={2}]{Jingguang}{Tian}
\author[affiliation={1}]{Jiaming}{Zhou}
\author[affiliation={1}]{Hui}{Wang}
\author[affiliation={1}]{Jiabei}{He}
\author[affiliation={1}]{Shiwan}{Zhao}
\author[affiliation={3}]{\\Xiangyu}{Kong}
\author[affiliation={2}]{Desheng}{Hu}
\author[affiliation={2}]{Xinkang}{Xu}
\author[affiliation={2}]{Xinhui}{Hu}
\author[affiliation={1,*}]{Yong}{Qin}

\affiliation{TMCC, College of Computer Science}{Nankai University}{Tianjin, China}
\affiliation{Hithink RoyalFlush AI Research Institute}{Zhejiang}{China}
\affiliation{University of Exeter}{Exeter}{United Kingdom}
\email{sunhaoqin@mail.nankai.edu.cn}

\usepackage{comment}
\usepackage{multirow}
\begin{document}

\maketitle

\begin{abstract}
     The Contrastive Language-Audio Pretraining (CLAP) model has demonstrated excellent performance in general audio description-related tasks, such as audio retrieval. However, in the emerging field of emotional speaking style description (ESSD), cross-modal contrastive pretraining remains largely unexplored. In this paper, we propose a novel speech retrieval task called emotional speaking style retrieval (ESSR), and ESS-CLAP, an emotional speaking style CLAP model tailored for learning relationship between speech and natural language descriptions. In addition, we further propose relation-augmented CLAP (RA-CLAP) to address the limitation of traditional methods that assume a strict binary relationship between caption and audio. The model leverages self-distillation to learn the potential local matching relationships between speech and descriptions, thereby enhancing generalization ability. The experimental results validate the effectiveness of RA-CLAP, providing valuable reference in ESSD.
\end{abstract}
 
\section{Introduction}
The language-based speech retrieval task refers to the process of searching for speech recordings based on natural language descriptions. This approach enables users to intuitively retrieve speech based on desired voice characteristics (e.g., emotion, speaker timbre, and speaking style) without relying on predefined labels. Different from general audio retrieval tasks, the natural language description required for speech retrieval focuses more on the paralinguistic and non-linguistic features contained in speech, such as emotion~\cite{liu23b_interspeech,sun2024iterative,sun2024fine,wang2025enhancing,sun2025enhancing} and speaker characteristics~\cite{he2025emotion}. By developing a speech retrieval system, it can be applied to various speech-related tasks, such as speech emotion captioning (SEC), speaking style captioning (SSC), and controllable Text-to-Speech (TTS).

In mainstream audio retrieval systems, architectures based on cross-modal contrastive learning have become dominant. These systems employ a dual-encoder framework to project natural language descriptions and audio recordings into a shared multimodal representation space, where contrastive learning is used to optimize the representations. Then, audio recordings are ranked based on their distance to the query descriptions. Recently, researchers have begun to explore the application of cross-modal contrastive learning in speech retrieval. For example, liu et al.~\cite{liu2023speaker} propose a learning framework based on contrastive learning that incorporates speaker labels into the learning process. Jing et al.~\cite{jing2024paraclap} utilize emotion labels and also incorporate acoustic and rhythmic features extracted by experts to generate queries and propose the ParaCLAP model. The above work is an initial exploration of speech retrieval tasks, which constructs emotion descriptions, speaker descriptions for speaker retrieval and emotion recognition respectively. However, whether using emotional attributes or speaker characteristics, such a simple and limited description faces inherent constraints in capturing the full complexity of speech attributes.

Similar natural language descriptions have emerged in more promising related fields. For example,
Guo et al.~\cite{guo2023prompttts} propose a PromptTTS system , which divides input prompts into style description and content description, exploring the possibility of using text descriptions (represented as prompts) to guide speech synthesis. Based on PromptTTS, Shimizu et al.~\cite{shimizu2024prompttts++} further propose the PromptTTS++ system, which divides the input prompts into style description and speaker description in a more detailed way. However, these controllable TTS systems ignore emotion attributes and may limit the expressiveness and naturalness of synthesized speech. Meanwhile, the limitations of SER have prompted researchers to turn to SEC to address the problem of modeling emotional expression. Researchers believe that emotions in speech are often multifaceted, and categorizing speech into a single emotional class is not sufficient to capture the complexity of emotional expression. Xu et al. introduce the SEC task and develop the SEcap framework, which extracts emotional features to generate corresponding emotional descriptions. Nevertheless, SEC focuses solely on emotion descriptions, without considering speaking style and speaker characteristics. Researchers are increasingly integrating these dimensions into their studies to provide a more comprehensive understanding of speech. Yamauchi et al.~\cite{yamauchi2024stylecap} propose StyleCap to automatically generate natural language descriptions of speaker, speaking style and emotion. 

\begin{figure*}[t]
  \centering
  \includegraphics[width=6.0in]{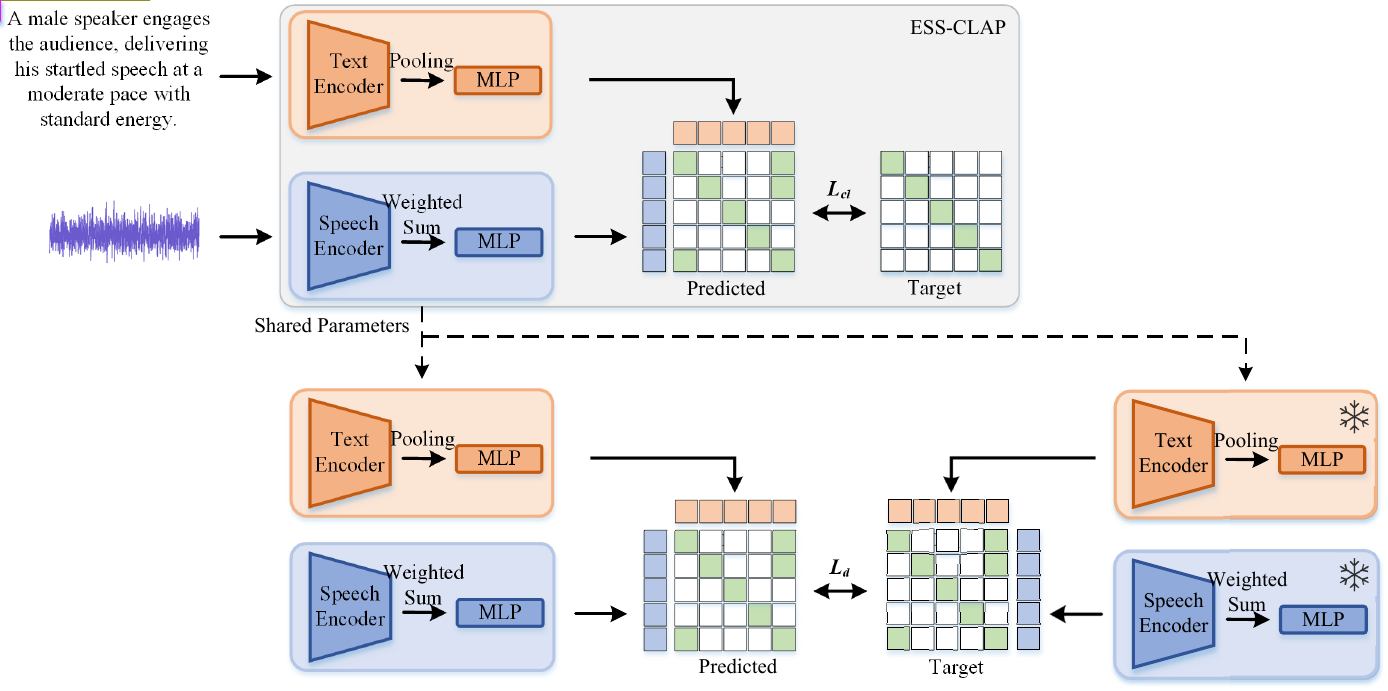}
  \caption{Overview of the proposed RA-CLAP model.}
  \label{fig:framework2}
\end{figure*}

We can observe that the basic goal of the mentioned work is to learn ideal speech and text representations to enable effective matching between speech attributes and text expressions. However, these findings have not yet been extended to the broader ESSR task. Inspired by contrastive pretraining~\cite{radford2021learning,wu2023large,pan2024gemo}, we propose the ESS-CLAP model, a contrastive language-audio pretraining method for emotional speaking style retrieval. Furthermore, since traditional methods assume a strict binary relationship between captions and audio segments, ignoring the possibility of partial matches, we introduce relation-augmented ESS-CLAP (RA-CLAP), which incorporates a two-stage training process. In the first stage, the ESS-CLAP model is trained. In the second stage, we use the pretrained model as a teacher model and apply self-distillation to learn the potential local matching relationships between speech and descriptions, thereby enhancing the model's generalization ability. Extensive experiments on three open-source datasets—PromptSpeech, TextrolSpeech, and SpeechCraft—are conducted to validate the usability and effectiveness of the ESS-CLAP model.

To the best of our knowledge, this study marks the first systematic application of contrastive cross-modal pretraining for the task of emotional speaking style retrieval. We hope that this work serves as a foundation for large-scale speech generative models applicable to tasks such as emotional speaking style captioning or prompt text-to-speech.

\section{Proposed Methods}

As shown in Fig.~\ref{fig:framework2}, the proposed RA-CLAP model is designed to learn a joint representation of speech and text through a contrastive learning and self-distillation framework. The training process consists of a pre-training phase and a self-distillation phase. The process is as follows: 

\subsection{pre-training phase}

Given a data pair consisting of a speech sample and its corresponding text description $(\mathbf{x}_i^{\text{speech}}, \mathbf{x}_i^{\text{text}})$, the ESS-CLAP model aims to project both modalities into a shared multimodal embedding space. 

\textbf{Speech Encoder:} A convolutional neural network combined with a transformer-based architecture (e.g., Wav2vec2.0, Hubert, and WavLM) is used to encode speech features into fixed-dimensional embeddings. Meanwhile, we also select two expert models Ecapa-TDNN and Resnet-34 for extracting speaker representations. 

To extract high-quality speech features, the SS-CLAP model utilizes the speech encoder, which processes raw speech waveforms and produces contextualized hidden representations. The speech encoder outputs a sequence of hidden states $\{\mathbf{h}_t\}_{t=1}^{T}$:
\begin{equation}
\{\mathbf{h}_t\}_{t=1}^{T} = \text{Encoder}_s(\mathbf{x}_i^{\text{speech}}),
\end{equation}
where $T$ denotes the number of time steps. Since different layers of the speech encoder capture various levels of acoustic and semantic information, we apply a weighted sum mechanism to aggregate these layer-wise features. Let $\alpha_l$ denote the learnable weight assigned to the $l$-th layer, and $\mathbf{h}_t^{(l)}$ represent the hidden state at time step $t$ from the $l$-th layer. The aggregated feature $\mathbf{z}_i^{\text{speech}}$ is computed as:
\begin{equation}
\mathbf{z}_i^{\text{speech}} = \sum_{l=1}^{L} \alpha_l \cdot \frac{1}{T} \sum_{t=1}^{T} \mathbf{h}_t^{(l)}.
\end{equation}
Here, $L$ denotes the total number of layers in the speech encoder, and the layer weights $\{\alpha_l\}$ are optimized during training. This weighted sum enables the model to focus on the most relevant layer-specific features, improving the alignment between the speech and text representations.

\textbf{Text Encoder:} To extract meaningful textual representations, the ESS-CLAP model employs a transformer-based text encoder (e.g., BERT or RoBERTa) to process the input text descriptions. Similar to speech encoder part, we use $\mathbf{z}_i^{\text{text}}$ to represent the high-level semantic representation of the text.

\begin{table*}[t]
    \caption{Performance comparison of different models on PromptSpeech and TextrolSpeech datasets (measured by mAP@10).}
    \centering
    \label{tab:map}
    \begin{tabular}{lcccc}
        \toprule
        Model & \multicolumn{2}{c}{PromptSpeech (mAP@10)} & \multicolumn{2}{c}{TextrolSpeech (mAP@10)} \\
        \cmidrule(lr){2-3} \cmidrule(lr){4-5}
        & Audio-to-Text & Text-to-Audio  & Audio-to-Text & Text-to-Audio  \\
        \midrule
        Wav2vec2.0 + BERT &8.5 & 9.6 &34.9 &39.7  \\
        Wav2vec2.0 + RoBERTa & 8.2&9.8& 37.4 &39.3 \\
        Hubert + BERT & 7.6& 8.8&35.1 &40.2 \\
        Hubert + RoBERTa & 8.0 &9.4& 37.0&40.8  \\
        WavLM + BERT & 14.6 &15.6&39.8&43.3\\
        WavLM + RoBERTa (ESS-CLAP) & \textbf{14.7} & \textbf{16.5} & \textbf{39.9}& \textbf{43.5}\\
        \midrule
        RoBERTa + Ecapa-TDNN & 7.2&8.6&37.4&38.5  \\
        RoBERTa + Resnet-34 &  7.6 &9.4 &35.2 &36.9 \\
        \midrule
        \bottomrule
    \end{tabular}
\end{table*}

\begin{table*}[t]
    \small
    \centering
    \caption{Comparison of retrieval performance (R@1, R@5, R@10) for Text-to-Audio and Audio-to-Text tasks on PromptSpeech and TextrolSpeech datasets. $^*$ represents ESS-CLAP model trained with a balanced sampling strategy. $^{**}$ represents RA-CLAP model.}
    \label{tab:R}
    \begin{tabular}{cccccccccccccc}
        \toprule
          \multirow{4}{*}{Training} &\multirow{4}{*}{Fine-tuning}  & \multicolumn{6}{c}{PromptSpeech} & \multicolumn{6}{c}{TextrolSpeech} \\
        \cmidrule(lr){3-8} \cmidrule(lr){9-14}
        & & \multicolumn{3}{c}{Audio-to-Text} & \multicolumn{3}{c}{Text-to-Audio} & \multicolumn{3}{c}{Audio-to-Text} & \multicolumn{3}{c}{Text-to-Audio} \\
        \cmidrule(lr){3-5} \cmidrule(lr){6-8} \cmidrule(lr){9-11} \cmidrule(lr){12-14}
        & & R@1 & R@5 & R@10 & R@1 & R@5 & R@10 & R@1 & R@5 & R@10 & R@1 & R@5 & R@10 \\
        \midrule
          PS &- &6.3 & 24.4 & 42.8 & 6.2 & 28.5  & 47.0 & - & - & - & - & - & -  \\
          TS &- &- & - & - & - & - & - & 23.5 & 63.5 & 82.0& \textbf{28.0}& 65.5& \textbf{87.5}\\
          SC &- &0.5 & 2.7 & 4.4 & 0.6&2.2&3.5 & 2.5&9.5&16.0&2.5&11.0&17.5  \\
          PS+TS &-& 4.9 & \textbf{27.8} & \textbf{45.8} & \textbf{6.8} & 28.7 & 48.8 & 23.5 & 62.0 & 87.0 & 26.0 & 65.5 & 86.5 \\
          PS+TS+SC &-& 1.0&4.6&9.5&2.1&6.7&11.3&1.5&11.0&21.0&3.5&13.5&21.0 \\
          PS+TS+SC &PS & \textbf{6.4} & 27.1 & 44.7 & 6.0 & \textbf{28.8}  & \textbf{49.4} & - & - & - & - & - & -  \\
          PS+TS+SC & TS & - & - & - & - & - & - & \textbf{26.5} & \textbf{66.0} & \textbf{89.0}& 27.5& \textbf{70.0}& 87.0\\
        \midrule
        PS+TS+SC$^*$ &-& 6.2&26.4&45.8&5.9&27.4&47.2&23.5&59.0&82.0&25.5&64.5&81.0 \\
        PS+TS+SC$^{**}$ &PS+TS&5.5&23.4&41.7&5.4&26.9&45.5&26.0&61.0&80.0&27.5&65.0&84.0\\
        \bottomrule
    \end{tabular}
\end{table*}

\textbf{Projection Layers:} The outputs of both encoders are passed through projection layers to map them into a shared latent space. It consists of two fully-connected layers and a RELU activation function, where the last fully-connected layer maps the representations to 512 dimensions.

\textbf{Contrastive Loss:} The InfoNCE loss is employed to train the model. For a given batch, positive pairs (matching speech and text) are encouraged to have high similarity, while negative pairs are pushed apart. The loss is computed as:
\begin{equation}
\mathcal{L}_{\text{cl}} = -\frac{1}{N} \sum_{i=1}^{N} \log \frac{\exp(\text{sim}(\mathbf{z}_i^{\text{speech}}, \mathbf{z}_i^{\text{text}}) / \tau)}{\sum_{j=1}^{N} \exp(\text{sim}(\mathbf{z}_i^{\text{speech}}, \mathbf{z}_j^{\text{text}}) / \tau)}
\end{equation}
where $\mathbf{z}_i^{\text{speech}}$ and $\mathbf{z}_i^{\text{text}}$ are the projected speech and text embeddings. $\tau$ is a temperature parameter, and \textit{sim} denotes cosine similarity.

\subsection{Self-Distillation}
In this stage, we utilize the pretrained model ESS-CLAP as the teacher model, freezing its parameters and initializing the student model with these parameters.

To overcome the limitations of traditional binary correspondence, we use the probability distribution predicted by the teacher model as the learning objective. By modeling the probability distribution, we could more accurately capture the relationship between speech and descriptions, thereby improving retrieval performance. The training loss $L_d$ is calculated as follows:
\begin{align}
C^{s} = \epsilon_{s} \times (\mathbf{z}_i^{\text{speech}} \cdot (\mathbf{z}_i^{\text{text}})^{T}),\\
C^{t} = \epsilon_{t} \times (\mathbf{z}_i^{\text{text}} \cdot (\mathbf{z}_i^{\text{speech}})^{T}),
\end{align}
where $C^{s}$ and $C^{t}$ are the similarity matrices of the RA-CLAP outputs. $\epsilon_{s}$ and $\epsilon_{t}$ are temperature hyper-parameters.

Subsequently, guided by the probability distribution matrix $M$ of the ESS-CLAP output, we use the KL loss to train our proposed RA-CALP:
\begin{align}
L_d = \frac{1}{2} \left( \text{KL}(ls(C^{s}), s(M)) + \text{KL}(ls(C^{t}), s(M)) \right),\\
\text{KL}(P \parallel Q) = \sum_{i,j} P(i,j) \log \frac{P(i,j)}{Q(i,j)},
\end{align}
where $s(\dots)$ represents softmax function and $ls(\dots)$ represents log\_softmax function. $i$ and $j$ represent the indexes of the row and column of the two-dimensional matrix.

\begin{table}[t]
    \centering
    \caption{Dataset Statistics for Speech Data Sources.}
    \label{tab:datasets}
    \begin{tabular}{lccc}
        \toprule
        \textbf{Dataset} & \textbf{Language} & \textbf{\#Duration} & \textbf{\#Samples} \\
        \midrule
        \textsc{SpeechCraft} & En & 1437.57h & 1,097,989 \\
        \textsc{PromptSpeech} & En & 38h & 26,588 \\
        \textsc{TextrolSpeech} & En & 330h & 236,220 \\
        \bottomrule
    \end{tabular}
\end{table}

\section{Experiments}
\subsection{Datasets}
The PromptSpeech (PS) dataset~\cite{guo2023prompttts} consists of prompts containing style and content information, along with corresponding speech samples. The dataset is available in two versions: a synthetic version and a real version. The synthetic version contains speech generated using a commercial TTS API, while the real version is derived from the LibriTTS~\cite{zen2019libritts} dataset.

The TextrolSpeech (TS) dataset~\cite{ji2024textrolspeech}, developed by Zhejiang University, provides 330 hours of speech data annotated with 236,220 natural text descriptions. The speech samples are sourced from multiple publicly available datasets (e.g., LibriTTS~\cite{zen2019libritts} and VCTK~\cite{yamagishi2019cstr}) as well as emotional datasets, including ESD~\cite{zhou2021seen}, TESS~\cite{dupuis2010toronto}, MEAD~\cite{wang2020mead}, SAVEE~\cite{haq2009speaker}, and MESS~\cite{morgan2019categorical}.

\begin{table*}[t]
    \small
    \centering
    \caption{Examples of the descriptions in different corpora.}
    \label{tab:example}
    \begin{tabular}{p{5cm}p{5cm}p{5cm}}
        \hline
        \textbf{PromptSpeech} & \textbf{TextrolSpeech} & \textbf{SpeechCraft} \\ \hline
        Her sound height is really high, the volume is normal, but she speaks very slowly.&A male speaker engages the audience, delivering his startled speech at a moderate pace with standard energy.&With a natural emotion, a young female with a normal pitch and low volume speaks slowly, expressing her emotions naturally, saying, in a calm tone.  \\ \hline
    \end{tabular}
\end{table*}

The SpeechCraft (SC) dataset~\cite{jin2024speechcraft} comprises 2,000 hours of bilingual speech and over 2 million annotated segments with detailed natural language descriptions. The dataset is sourced from multiple publicly available speech corpora, including AISHELL-3~\cite{shi2020aishell} and Zhvoice\footnote{https://github.com/fighting41love/zhvoice} for Chinese, as well as GigasSpeech-m~\cite{chen2021gigaspeech} and LibriTTS-R~\cite{koizumi2023libritts} for English. In this paper, only the English data are used.

\subsection{Experimental Setup}
In all experiments, we employ the Adam optimizer with hyperparameters $\beta_1$ = 0.9 and $\beta_2$ = 0.99 to optimize the proposed CLAP-based methods. The learning rate is set to 2e-5, and the batch size is fixed at 192. All models are trained for 15 epochs. To address the issue of imbalanced learning due to significant differences in sample size across the three datasets, we use a balanced sampling strategy. In each training batch, we ensure that the sample ratio from the three datasets remains 1:1:1.

\subsection{Text-to-Audio Retrieval}

\textbf{speech and Text Encoders:} we first perform experiments on retrieval tasks to identify the optimal combination of text and speech encoders. Specifically, we systematically evaluate several text encoders (e.g., BERT, RoBERTa) and speech encoders (e.g., Wav2vec2.0, HuBERT, WavLM). For each encoder pair, we conduct both text-to-audio retrieval and audio-to-text retrieval tasks. The performance of each combination is assessed using the mean average precision at top 10 (mAP@10), which measures the average accuracy of relevant items within the first 10 retrieval results.

The table~\ref{tab:map} presents a performance comparison of six model combinations on the PromptSpeech and TextrolSpeech datasets. Overall, ESS-CLAP achieved the best results, obtaining the highest scores for both Audio-to-Text and Text-to-Audio on the PromptSpeech and TextrolSpeech datasets (14.7/16.5 and 39.9/43.5, respectively). For speech encoders, WavLM outperforms Wav2vec2.0 and Hubert, while for text encoders, RoBERTa demonstrates superior performance in text modeling compared to BERT. Additionally, since speaker representations provide personalized timbre characteristics that help models understand how different speakers express emotionally, we conduct experiments on two mainstream architectures, ECAPA-TDNN and ResNet-34. The results show that speaker representations achieve comparable performance to deep representations extracted by SSL models, highlighting their potential in modeling emotional speaking style.

\textbf{Dataset Scale:}
We adopt WavLM-RoBERTa as the optimal model setup for the retrieval experiments. Table~\ref{tab:R} shows the retrieval performance of the models trained on PS, TS, SC, and combined datasets (PS + TS and PS + TS + SC) respectively.

Firstly, models trained on a single dataset perform better on their corresponding test sets. Specifically, the model trained on the PS dataset achieves R@10 scores of 42.8\% and 47.0\% for the Audio-to-Text and Text-to-Audio tasks, respectively. Meanwhile, the model trained on the TS dataset attains significantly higher R@10 scores of 82.0\% and 87.5\%, outperforming other training strategies. In contrast, the model trained on the SC performs poorly in both datasets. Based on the data volume and description style presented in Table \ref{tab:datasets} and Table \ref{tab:example} , it can be inferred that the SC dataset features more flexible and diverse descriptions, increasing the difficulty of model learning. Additionally, since the SC dataset is significantly larger than the PS and TS datasets, the model tends to learn the SC-style descriptions, which negatively impacts its retrieval performance on other datasets. 

Secondly, when the model is trained on the PS + TS, its retrieval performance improves across both datasets, indicating that joint training helps enhance the model’s performance in cross-dataset tasks. However, when the SC dataset is introduced further, the performance of the model drops significantly. This result indicates that the inclusion of the SC dataset negatively impacts the model's ability to learn from other datasets, supporting the earlier hypothesis that SC data is more challenging to learn and that its description style affects the adaptability of the model. 

Moreover, after further fine-tuning the pretrained model on PS and TS, the model's performance improves. It is worth noting that although the model trained only on the SC dataset performs poorly, the large-scale expansion of the SC dataset still holds potential to enhance the model's capabilities. A larger dataset helps enrich the model's representational power and improves its adaptability to diverse descriptions, ultimately boosting retrieval performance after appropriate fine-tuning. Therefore, while the inclusion of SC data without fine-tuning may lead to performance degradation, the rich information it provides can still effectively improve the model's generalization ability and cross-dataset adaptability within a joint training and fine-tuning framework.

Finally, we compare the retrieval performance of ESS-CLAP and RA-CLAP with data balanced sampling strategy. The results show that this strategy significantly improves retrieval performance and partially mitigates the issues caused by the large proportion of SC and the considerable differences in descriptions. We also find that after self-distillation fine-tuning, RA-CLAP does not improve performance on the PS evaluation set, but achieves better performance on the TS evaluation set. One reason for this is that the PS dataset does not include emotion descriptions. When the model learns emotion descriptions from the TS dataset, the learned distribution deviates from the PS dataset, indicating that the model makes a trade-off between different types of descriptions to maintain its adaptability to diverse descriptions.

\section{Conclusions}
In this paper, we conduct a preliminary investigation into the key factors affecting the CLAP model for ESSR task. As part of this study, we propose a novel task ESSR and ESS-CLAP, a CLAP model specifically designed for learning speech representations by effectively integrating speech with natural language descriptions. Further, we propose the RA-CLAP, which re-learns potential part-matching relationships between speech and description by self-distillation. Overall, this study marks the first systematic application of contrastive cross-modal pretraining in emotional speaking style description, demonstrating its significant potential and laying a solid foundation for tasks such as emotional speaking style caption and promptTTS.

\bibliographystyle{IEEEtran}
\bibliography{mybib}

\end{document}